\def\rco {$^{12}R_{32}$}
\def\rhi {$^{iso}R_3$}
\def\rlo {$^{iso}R_2$}
\def\hii {\ion{H}{2}}
\def\hi  {\ion{H}{1}}
\def\mion {$\dot M_{ionized}$}
\def\msunyr {\hbox{$M_\odot$ yr$^{-1}$}}
\def\cmc {cm$^{-2}$}
\def\mdot {$\dot M$}
\def\jj#1#2{\hbox{$J$ = #1$\rightarrow$#2}}
\def\refindent{\advance\leftskip by 24pt\parindent=-24pt}
\def\figcap#1#2{{\refindent \tup{F}{IG}. {#1}.--- {#2}\par}}
\def\samename{\vrule height0.4pt depth0.0pt width1.0in \thinspace.}
\def\tabnormal{\normalsize}
\def\tabsmall{\scriptsize}
\def\tabbaseskip{\baselineskip=14pt}
\def\tablesize{\tabnormal\tighten\parskip=0pt\tabbaseskip}
\def\tup#1#2{{\uppercase{\tabnormal{#1}\tabsmall{#2}}}}
\def\note{\tup{N}{OTE}.---}

\def\refs{\tup{R}{EFERENCES}.---}
\def\linefil{\leftskip=0pt plus0pt\rightskip=0pt plus0pt}
\def\goodfil{\linefil\spaceskip=3.3333pt plus 3.3333pt}
\def\na{$\cdots$}
\def\tabvs{\noalign{\vskip 5pt}}
\documentstyle[aaspp]{article}
\begin{document}
\title{Extremely High Velocity Outflows}
\author{Minho Choi\altaffilmark{1},
        Neal J. Evans II\altaffilmark{2},
        and Daniel T. Jaffe\altaffilmark{3}}
\affil{Department of Astronomy, University of Texas at Austin,
       Austin, Texas 78712--1083}
\altaffiltext{1}{Electronic mail: minho@astro.as.utexas.edu}
\altaffiltext{2}{Electronic mail: nje@astro.as.utexas.edu}
\altaffiltext{3}{Electronic mail: dtj@astro.as.utexas.edu}

\begin{abstract}

Extremely high velocity (EHV) wings,
with full widths of 72 to 140 {km~s$^{-1}$},
are seen on the CO \jj32\ lines
toward W3 IRS 5, GL 490, NGC 2071, W28~A2 \hbox{(G05.89--0.39)},
GL~2591, S140, and Cepheus A.
Observations of $^{12}$CO and $^{13}$CO \jj32\ and \jj21\ lines indicate
that optical depth generally decreases
with increasing velocity separation from the ambient cloud velocity.
Maps of the extremely high velocity
($|V-V_0|$
{$\rlap{\raise.4ex\hbox{$>$}}\lower.55ex\hbox{$\sim$}\,$}\ 20 {km~s$^{-1}$})
and the high-velocity
\hbox{(5 {$\rlap{\raise.4ex\hbox{$<$}}\lower.55ex\hbox{$\sim$}\,$}\ $|V-V_0|$
{$\rlap{\raise.4ex\hbox{$<$}}\lower.55ex\hbox{$\sim$}\,$}\ 20 {km~s$^{-1}$})}
CO emission components show
that the morphology of the two components is similar
in W3 IRS 5 and W28~A2 but may be different in GL~2591, S140, and Cepheus A.

The results of our survey suggest
that EHV wings are common around infrared sources
of moderate to high luminosity [500 to (4{${} \times 10^{5}$}) {$L_\odot$}]
in dense regions.
Line ratios imply that the EHV gas is usually optically thin and warm.
Characteristic velocities range from 20 to 40 {km~s$^{-1}$},
yielding timescales of \hbox{1600--4200 yr}.
Since most sources in this study are producing some ionizing photons,
these short timescales suggest
that neutral winds coexist with ionizing photons.

We examined two possible sources for the extremely high velocity CO emission:
a neutral stellar wind; and swept-up or entrained molecular gas.
Neither can be ruled out.
If the high-velocity (HV) gas is swept up
by a momentum-conserving stellar wind
traced by the extremely high velocity CO emission,
most of the C in the winds from luminous objects cannot be in CO.
If the EHV and HV forces are equal,
the fraction of C in a form other than CO
increases with source luminosity
and with the production rate of ionizing photons.
This trend is natural in the stellar wind hypothesis,
but models of winds around such luminous objects are needed.
We consider other possible chemical states
for the carbon in the stellar wind.

\end{abstract}


\clearpage
\section {INTRODUCTION}

Molecular outflows are common in regions of star formation
(about 200 known:
Lada 1985; Fukui et~al. 1993; Bachiller \& G\'omez-Gonz\'alez 1992).
High-velocity (hereafter HV) molecular flows---with
velocities ranging from a few {km~s$^{-1}$}\ to about 20 {km~s$^{-1}$}---have
been studied extensively
(Bally \& Lada 1983; Plambeck, Snell, \& Loren 1983; Snell et~al. 1984;
Richardson et al. 1985; Phillips et al. 1988).
These molecular outflows most likely consist of ambient cloud material
swept up by a faster stellar wind (Snell, Loren, \& Plambeck 1980).

Searches for an ionized fast stellar wind to drive the HV gas found
that the momentum in the ionized gas was inadequate
to sweep up the observed amount of molecular material
(Rodr{\'\i}guez \& Cant\'o 1983; Evans et al. 1987),
leading to the suggestion that the stellar wind was primarily neutral
(Natta et al. 1988). The neutral wind gained support from the
discovery of very wide wings on the spectra of \hi\ emission
(Lizano et al. 1988; Giovanardi et~al. 1992)
toward several young stars.
Lizano et~al. (1988), Koo (1989, 1990), and Margulis \& Snell (1989)
also recognized extremely high velocity (hereafter EHV) wings on the CO spectra
toward a few sources, suggesting that there might be CO
in the stellar wind itself.
In recent years, there have been
additional observations of EHV CO outflows
(Bachiller \& Cernicharo 1990; Masson, Mundy, \& Keene 1990;
Bachiller, Mart{\'\i}n-Pintado, \& Planesas 1991; Mitchell \& Hasegawa 1991;
Chernin \& Masson 1992; Richer, Hills, \& Padman 1992).
Stahler (1993) has suggested that EHV outflows are the manifestation in
young objects of the optical jets seen in later phases of evolution.

Calculations of the chemistry
in stellar winds from low-mass stars indicated
that CO could form in the neutral wind
(Glassgold, Mamon, \& Huggins 1989).
Later calculations of the thermal structure of these winds
(Ruden, Glassgold, \& Shu 1990) agreed with that conclusion,
and new chemical models which incorporated the thermal structure found
that, for mass-loss rates
\hbox{\mdot\ $\geq$ 3{${} \times 10^{{-6}}$}\ \msunyr}
virtually all of the carbon is in CO
([CO]/[H] = \hbox{4{${} \times 10^{{-4}}$}}),
even though the hydrogen remains mostly atomic,
unless \mdot\ $\geq$ 10$^{-4}$ \msunyr
(Glassgold, Mamon, \& Huggins 1991).

The thermal models of Ruden et al. (1990) raised questions, however,
about whether a wind could maintain
the high temperature \hbox{($T_K \approx$ 100 K)}
and density \hbox{($n > 10^4$ {cm$^{-3}$})}
deduced by Koo (1989) for \hbox{HH 7--11},
on large enough scales ($\sim$0.1~pc)
to be consistent with the observations.
In addition, Masson et~al. (1990) and Bachiller \& Cernicharo (1990)
found relatively narrow ($\Delta V \approx$ \hbox{20 {km~s$^{-1}$}})
spectral features at extremely high velocities
($|V-V_0| \approx$ \hbox{150 {km~s$^{-1}$}}),
displaced from the center of the outflow in \hbox{HH 7--11},
suggesting that the EHV gas consists of discrete blobs,
rather than a continuous wind.
Similar discrete EHV spectral features exist
in several other sources
(Bachiller et al. 1990, 1991; Bachiller \& G\'omez-Gonz\'alez 1992).
The blobs may be related to Herbig-Haro objects
and to localized EHV spectral features
often found in H$_2$O maser observations of compact \hii\ regions
(Genzel \& Downes 1977, 1979).

The nature of the EHV CO emission is currently unclear. It could
arise in the stellar wind, or it could be ambient
molecular gas which has been entrained or swept up (see Bachiller \&
G\'omez-Gonz\'alez 1992 and Stahler 1993 for excellent reviews).

While observing the CS $J$ = 7{$\rightarrow$}6 line
toward several sources, we noticed EHV wings on the CO \jj32\ lines,
which appear in the opposite sideband.
We followed up by mapping the EHV gas in CO $J$ = 3{$\rightarrow$}2 emission
and by obtaining data on other transitions.
In this paper, we report observations of W3~IRS~5, GL~490, NGC~2071, W28~A2,
GL~2591, S140, and Cepheus~A.
Most previous detections of EHV wings have been
toward sources of low to moderate luminosity ($L< 1000$ {$L_\odot$});
this work extends the phenomenon to higher luminosity sources.
These sources, some of which are generating ionizing photons,
may have different wind properties if ionization plays a role in the wind
generation process or in the chemical composition of the wind.

We describe our observations in \S\ 2.
Section 3 explains the analysis
we used to get physical parameters from the observations
(optical depth, excitation temperature, mass, etc.).
We summarize our analysis and discuss its implications in \S\ 4.

\section {OBSERVATIONS AND DATA REDUCTION}

We made all observations with the 10.4 m telescope
of the Caltech Submillimeter Observatory
(CSO)\footnote{The CSO is operated
by the California Institute of Technology
under funding from the National Science Foundation, contract AST 90--15755.}
at Mauna Kea during several runs between 1989 June and 1992 February.
At the central position,
all sources were observed in four molecular transitions:
CO \jj32, CO \jj21,
$^{13}$CO \jj32, and $^{13}$CO \jj21.
Since most of the sources show much brighter CO $J$ = 3{$\rightarrow$}2 wings
than CO \hbox{$J$ = 2{$\rightarrow$}1} wings,
they were mapped in the CO $J$ = 3{$\rightarrow$}2 line.
Table 1 summarizes the observational parameters.
The data were obtained using the 230~GHz and 345 GHz SIS receivers
in double-sideband mode
with a 500~MHz AOS spectrometer as the backend.
Telescope pointing was checked by observing CRL~618, CRL 2688,
or Jupiter regularly and was consistent to within $\sim$5{${''}$}.

The data were taken by position switching.
Table 2 lists the central and the {\scriptsize OFF} positions.
For each source, an effort was made to acquire at least some of the data
with an {\scriptsize OFF} position
which was free of CO emission in the line core.
Since some of these {\scriptsize OFF} positions were quite distant,
other data were taken with more nearby {\scriptsize OFF} positions
to improve the baselines.
Therefore, the line cores in the figures may be affected by emission
in the {\scriptsize OFF} positions.
Some sources were observed repeatedly over several runs
and the intensities were consistent to within $\sim$10\%.

The antenna temperature ($T^*_A$) was calibrated
by the standard chopper-wheel method, which automatically corrected,
to first order, for the effects of atmospheric attenuation.
The radiation temperature, $T_R$, following the notation
of Kutner \& Ulich (1981), was obtained by dividing $T^*_A$
by the telescope main beam efficiency
($\eta_{\rm mb} = \eta_b /\eta_{\rm rss}$,
where $\eta_b$ is the beam efficiency defined by Kraus (1966)
and $\eta_{\rm rss}$ is the rear spillover and scattering efficiency;
see Table~1).
We determined $\eta_b$ by observations of planets,
primarily Jupiter (Mangum 1993).
Thus, the calibration is appropriate for sources which fill the main beam.
For sources smaller than the main beam, filling factors must be considered.
Since the beam was about 30{${''}$}\ for the $J$ = 2{$\rightarrow$}1 data
and 20{${''}$}\ for the $J$ = 3{$\rightarrow$}2 data,
a point source centered in both beams will have
a filling factor for the $J$ = 3{$\rightarrow$}2 data
which is (3/2)$^2$ times as high as that for the $J$ = 2{$\rightarrow$}1 data.
For the CO $J$ = 3{$\rightarrow$}2 observations,
the CS \hbox{$J$ = 7{$\rightarrow$}6} appears in the image sideband.
The velocity of the telescope with respect to the LSR
was different on each run;
as a result, the intensities and shapes of the CS $J$ = 7{$\rightarrow$}6 lines
in Figures 1{\it a\/}--7{\it a\/} are not correct,
and in some spectra, the CS line appears twice.

The data were averaged for each line and a baseline was removed,
using the velocity ranges listed under {\it Baseline\/} in Table 3.
For the CO $J$ = 3{$\rightarrow$}2 line, the region of the CS \jj76\ line
was also excluded from the baseline fit.
At the central positions, where we take line ratios,
first order baselines were used.
The resulting RMS noise levels in the baselines are given in Table 3
and spectra are shown in Figures 1{\it a\/}--7{\it a\/}.
Both a full-scale plot and a blowup are shown to indicate the range
and quality of the baselines.
{}From the blown-up spectra, one can see
that the full widths to where the wing vanishes into the noise
range from 72~{km~s$^{-1}$}\ for W3~IRS~5
to 140 {km~s$^{-1}$}\ for W28~A2 (Table 4).

The mapping data were taken with shorter integration times,
so their quality is lower.
First-order baselines were also used for most of these spectra,
but a few off-center positions required baselines up to third order.

Since we are interested in the differences
between HV components and EHV components of outflows,
we have defined the inner and outer wings
by examination of the CO \jj32\ spectra.
For about half of our sources,
the boundary between the inner and outer wings was set
by noticing a distinct change in the CO \jj32\ line profile,
marked by the appearance of a component
which declines more slowly with velocity away from the line core.
For sources without a distinct change of slope,
we set the boundary between our inner and outer wings
at the largest velocity at which wings were previously known.
For W28~A2, the boundaries between inner and outer wings are quite subjective
because it is difficult to find a distinct change of slope
and there is no reference useful to define the boundary.
The average velocity offset from the line center
for the boundary between inner and outer wings was 21 {km~s$^{-1}$},
corresponding roughly to the boundaries
chosen in other studies (e.g., Koo 1990)
making it a reasonable, if not precise, choice
for separating HV and EHV emission.
The outer boundaries of the outer wings were set
by our subjective assessment of the extremes of detectable wings.
The wing boundaries are listed in Table 4
and marked in the blown-up spectra and the line ratio plots
by dashed vertical lines.

After defining the wing boundaries,
we integrated the intensity between the boundaries
and plotted the results as contour maps.
We always set the lowest contour level higher than twice the RMS noise.

To assess the optical depths in the wings,
we examined various ratios of lines as a function of velocity.
We resampled the spectra toward the central positions
to a resolution of 2 {km~s$^{-1}$}\ and formed the following line ratios:
\rco, CO \jj32\ line divided by CO \jj21;
\rhi, CO \jj32\ divided by $^{13}$CO \jj32;
and \rlo, CO \jj21\ divided by $^{13}$CO \jj21.
The results \hbox{(Figs. 1{\it c\/}--7{\it c\/})} were blanked
from the plot beyond the velocity
where either the numerator or the denominator fell below
the threshold (3 times the RMS noise in the resampled spectra) in Table 3.

\section {ANALYSIS}

In this section, we describe how we determined optical depths,
excitation temperatures, and column densities from the line ratios.
We also explain our calculations of the mass, momentum,
and kinetic energy in the outflowing gas. We then compare the driving
force in the EHV gas to that in the HV gas,
compute timescales for the outflows, and collect data on the stellar
luminosity and the
production rate of ionizing photons.

\subsection {Optical Depth}

{}From the isotopic line ratio, \rhi, we calculated $\tau _{32}$,
the optical depth of the CO \jj32\ line in each channel, from
\begin{equation}
^{\rm iso}R_3 = {{T_{R3-2}}\over {T_{R(13)3-2}}}
              = {{1 - \exp(-\tau _{32})}\over {1 - \exp(-\tau _{32}/X)}},
\end{equation}
where $X$ (listed in Table 5) is the isotopic abundance ratio,
[CO]/[$^{13}$CO].
We assumed that $X$ is a linear function of distance
from the Galactic center (Langer \& Penzias 1990) except for NGC~2071,
where we assumed $X$ = 73, the average for Ori~A.
The low strength
in the outer wings of $^{13}$CO lines means that
most sources do not have well-defined \rhi\ at extremely high velocities.

One way to constrain the optical depth at extremely high velocities is
through the ratio of the CO \jj32\ and $J$ = 2{$\rightarrow$}1 lines;
the ratio is given by
\begin{equation}
^{12}R_{32} = {{T_{R3-2}}\over {T_{R2-1}}}
   = \phi\
     {{[J_{\nu 32}(T_{\rm ex}) - J_{\nu 32}(T_{\rm bk})]}
      \over {[J_{\nu 21}(T_{\rm ex}) - J_{\nu 21}(T_{\rm bk})]}}\
     {{[1 - \exp(-\tau_{32})]}\over {[1 - \exp(-\tau_{21})]}},
\end{equation}
where $\phi$ is the ratio of beam filling factors
which accounts for differences in telescope beam sizes,
$J_\nu(T)$ is the Planck function in temperature units,
$T_{ex}$ is the excitation temperature
(assuming that the two lines have the same excitation temperature),
and \hbox{$T_{bk}$ = 2.7 K}.
In the limits of \hbox{$\tau$ {$\rightarrow$}\ 0}
and $T_{ex}$ {$\rightarrow$}\ $\infty$,
\rco\ approaches ($\nu_{32}$/$\nu_{21}$)$^2$ = (3/2)$^2$
(for an extended source) or ($\nu_{32}$/$\nu_{21}$)$^4$ = (3/2)$^4$
(for a point source centered in both beams)
as shown in Figure 10 (Appendix A).

Most of the sources have \rhi\ ratios
which increase away from the line center
and \rco\ ratios which imply low optical depth in the outer wings.
We assumed that the outer wings were optically thin,
except for the red wing of S140 and both wings of GL 490,
for which \rco\ is relatively low.
For those wings, we assumed $\tau_{32} = 2$ (which is chosen arbitrarily)
for all velocities.
For the inner wings, we fitted $\tau_{32}$, determined from \rhi,
with an exponential function ($\tau_{32} = \alpha \exp[\beta (V-V_0)]$,
where $\alpha$ and $\beta$ are constants determined by a least-squares fit)
for each side of the spectrum \hbox{(Figs. 1{\it c\/}--7{\it c\/})}.

\subsection {Excitation Temperature}

For most of the outer wings,
the optical depth is low and we can calculate $T_{ex}$
with \rco\ averaged over the wing intervals from
\begin{equation}
^{12}R_{32} = \phi\
              {{[J_{\nu 32}(T_{\rm ex}) - J_{\nu 32}(T_{\rm bk})]}
               \over {[J_{\nu 21}(T_{\rm ex}) - J_{\nu 21}(T_{\rm bk})]}}\
              {{\tau_{32}}\over {\tau_{21}}};
\end{equation}
assuming LTE,
\begin{equation}
{{\tau _{32}}\over {\tau _{21}}} = {3\over 2}\
   {{1 - \exp(-16.597/T_{\rm ex})}\over {\exp(11.065/T_{\rm ex}) - 1}}.
\end{equation}
It is hard to determine $\phi$,
but we can calculate a very firm lower limit to $T_{ex}$
by assuming a point source centered in both beams
[i.e., $\phi\ = (3/2)^2$].
Although some maps show extended sources,
the EHV wind may have a complex structure in position-velocity space
that does not allow us to use extended emission
to rule out a large value of $\phi$.

Table 5 shows the lower limits to $T_{ex}$ calculated from equation (3);
they range from 16 to 94~K,
indicating that at least some of the EHV gas is warm.
If $\phi$ is closer to unity, the temperatures must be much higher.
For example, if $\phi$ = 3/2, an intermediate value
between that expected for a point source and the value for an extended source,
$T_{ex}$ ranges from 30~K to $>$ 250 K.
Any mixture of cooler gas will produce a lower ratio; for example, applying
this method to mixtures of two temperature components,
with equal optical depth, would yield an average of the two temperatures.
The high observed ratios are thus a strong indication that the bulk of
the EHV gas is quite warm.

Equation (3) cannot be used to calculate the excitation temperature in the
inner wings because they are not, in general, optically thin.
An alternative procedure would be to calculate the excitation temperature
from the ratio of the two optical depths
deduced from \rhi\ and \rlo\
(for example, Snell et~al. 1984; Margulis \& Lada 1985).
While this method reduces systematic uncertainties due to beam filling,
it can seriously underestimate $T_{ex}$ if components with different optical
depths are present (Appendix~B).
Consequently, we have not estimated $T_{ex}$ in the inner wings, but we show
in the next section that knowledge of $T_{ex}$ is not essential for
determining column densities.

\subsection{Column Density}

The column density of CO for each channel, $N_i^{\rm CO}$ was calculated
from the intensity of the CO \jj32\ line from
\begin{equation}
N_i^{\rm CO} = 1.10\times 10^{15}\
               {{T_{R3-2}\ \Delta V} \over {D(n, T_K)}}\
               {{\tau _{32}}\over{1 - \exp(-\tau _{32})}},
\end{equation}
where
\begin{equation}
D(n, T_K) = f_2\ [J_\nu(T_{\rm ex})-J_\nu(T_{\rm bk})]\
            [1-\exp(-16.597/T_{\rm ex})],
\end{equation}
$\Delta V$ is the channel width in {km~s$^{-1}$},
and $f_2$ is the fraction of CO molecules in the \hbox{$J$ = 2} state.
$N_i^{\rm CO}$ is in units of cm$^{-2}$ and $\Delta V$ in {km~s$^{-1}$}.
The optical depth, $\tau_{32}$, was assigned as described in \S\ 3.1.
Assuming that the optical depth is known, the main uncertainty is
the value of the function $D(n, T_K)$.
If all the levels are not in LTE,
$D(n, T_K)$ depends on the density, as well as the temperature,
through the factor $f_2$.
The LTE assumption can lead to large overestimates of $N^{\rm CO}$
for high temperature and modest density.

Although $T_{ex}$ is not well known,
LVG simulations show that $D(n, T_K)$ does not vary much
within reasonable intervals of density, $n$,
and kinetic temperature, $T_{K}$ (Fig. 8).
Therefore, we assumed $D(n, T_K)$ = 1.5 in our calculations,
which is correct to less than a factor of 2
for $10 < T_K < 200$ K and $10^4 < n < 10^6$ cm$^{-3}$.
Column densities derived from \jj32\ line are somewhat less uncertain
than those derived from the \jj10\ line
because $D(n, T_K)$ varies from 0.074 to 0.76 for the \jj10\ transition
in the same intervals of $n$ and $T_K$.

\subsection{Mass per Channel: Two Possibilities}

The mass per channel, $M_i$, can be computed from the column density
per channel, $N_i^{\rm CO}$, the distance, $d$ (Table 2), and the solid angle
subtended by emission in that channel, $\Omega$, by
\begin{equation}
M_i = {\mu m_{\rm{H}} d^2 \Omega N_i^{\rm CO}}\
     {{[{\rm H}]}\over {[{\rm C}]}}\ {{[{\rm C}]}\over {[{\rm CO}]}},
\end{equation}
where $\mu = 1.3$, $m_{\rm H}$ is the mass of a hydrogen atom,
and [H]/[C] = 2.5{${} \times 10^{3}$}\ (Grevasse et~al. 1991).
The final factor, [C]/[CO], is about 8 in ambient molecular gas
(Dickman 1978; van Dishoeck et~al. 1992),
with the missing carbon presumably tied up in solid matter.
For stellar winds from low-mass young stars,
calculations indicate that molecule formation precedes dust grain formation,
so that all C is in CO for \mdot $> 10^{-4}$ \msunyr\
(Glassgold et al. 1991).
If correct, these models say
that a given $N_i^{\rm CO}$ would imply 8 times less matter
if the CO emission arises from a wind
than if it arises from swept-up gas.
Because the [C]/[CO] ratio in winds from more massive stars is not well known,
we will begin by assuming  that carbon is fully associated into CO.

In calculating mass, momentum, and energy in the next section, we
give two different results for the EHV gas (Table 5),
one for a stellar wind, with \hbox{[C]/[CO] = 1},
and one for swept-up gas, with \hbox{[C]/[CO] = 8}.
Since the HV gas is always assumed to be swept up,
we give only one result for it.

\subsection{Mass, Momentum, and Kinetic Energy}

The mass, momentum, and kinetic energy of each outflow were calculated from
\begin{equation}
M = \sum_i M_i,
\end{equation}
\begin{equation}
P = \sum_i M_i |V_i-V_0|,
\end{equation}
and
\begin{equation}
E_K = \sum_i {1\over 2} M_i (V_i-V_0)^2,
\end{equation}
with summations over the velocity channels, $i$, within each wing interval;
$V_0$ is the central velocity (see Table 4)
of the $^{13}$CO $J$ = 2{$\rightarrow$}1 line (W3 IRS 5 and Cepheus~A)
or $^{13}$CO \hbox{$J$ = 3{$\rightarrow$}2} line (all other sources).
Since we made no corrections to the velocities for projection effects,
the momenta and kinetic energies in Table 5 are lower limits.

Our estimates agree reasonably well with previous work.
The total momentum in the wings of W3~IRS~5 is about 80\%
that found by Mitchell, Hasegawa, \& Schella (1992),
who included some low velocities in their red wing
that we do not count as wing emission.
Because the GL~490 maps of Snell et~al. (1984) and Mitchell et al. (1992)
cover a larger area than ours,
their estimates of the mass, momentum, and kinetic energy
of the HV outflow from CO \jj21\ and \jj10\ observations
are about 3--4 times larger than ours.
The estimates of the masses, momenta, and kinetic energies of NGC 2071
by Chernin \& Masson (1992) are
larger (by a factor of about 5 for the outer wings
and a factor of about 2 for the inner wings),
mostly because their map is larger than ours.
The W28 A2 outflow is much more energetic than the others in this paper.
Our estimates of the momentum of the outflow
agree with Harvey \& Forveille (1988),
considering the slightly different definitions of the wings
and the fact that we do not include the inner red wing.
The blue outflows of GL 2591 are much more energetic and massive
than the red outflows.
Mitchell et al. (1992) also found
this asymmetry between the blue and the red outflows.
The mass of the combined red and blue HV outflows is
about twice the mass calculated by Mitchell et~al. (1992)
when their values are corrected to our distance.
Our estimates of the mass, momentum, and kinetic energy
of the HV outflows of S140
are consistent with those of Snell et al. (1984),
but exceed those of Hayashi et~al. (1987),
who did not see the full extent of even the HV wings.
Our estimate of the mass in the HV flow of Cepheus A
is about 2.5 times that of Ho, Moran, \& Rodr{\'\i}guez (1982).

\subsection{Driving Force}

In \S\ 4, we will consider the possibility that the EHV wing comes
from CO in the actual stellar wind. If the stellar wind drives the
HV flow, the average driving force in the wind must equal the average
force applied to the HV gas. The force is calculated from
\begin{equation}
F = {P\over t} = {{P^2}\over{M R}},
\end{equation}
where $t = R/V_{\rm ch}$ is the characteristic timescale for the outflow,
$R$ is the size
of the outflow, and $V_{\rm ch} = P/M$ is the characteristic (mass-weighted)
velocity.

The ratio of driving force in the EHV wing
to the average force required to sweep up the HV wing
was calculated from
\begin{equation}
R_F = {{F_{\rm EHV}}\over{F_{\rm HV}}} = {{R_P^2}\over{R_R\ R_M}},
\end{equation}
where $R_P$, $R_R$, and $R_M$ are
ratios of momentum, size, and mass, respectively.
It is hard to determine $R_R$ from the maps for most of our sources.
Since our maps and maps from previous work
(see references in Table 6)
show that the spatial extents of the EHV and the HV outflows are similar
(within a factor of $\sim$2),
we assumed $R_R = 1$ in our calculations.
This is not true for GL~2591,
which shows a very compact EHV source along with extended HV outflows.
So the correct $R_F$ of GL 2591
should be larger than the value listed in Table 6.
Because the correction factor for projection
should be similar for the HV and EHV outflows,
ratios of quantities for the EHV and HV outflows
should be unaffected by projection effects.

Table 6 shows $R_F$ for all of our sources
and for some EHV outflows observed by others.
Unlike our classification, the wings of HH 7--11 (Koo 1990)
and IRAS 03282+3035 (Bachiller et al. 1991) are
divided into HV, IHV (intermediate high velocity), and EHV wings.
For these two sources, we show two values in Table~6,
$(F_{EHV}+F_{IHV})/F_{HV}$ and $F_{EHV}/(F_{IHV}+F_{HV})$.
Also NGC~7538 IRS~9 (Mitchell \& Hasegawa 1991) has an EHV wing detected
only to the red side of the spectrum.
Again, we list two values in Table~6,
$F_{RedEHV}/F_{RedHV}$ and $F_{RedEHV}/(F_{RedHV}+F_{BlueHV})$.

Table 6 also shows the characteristic timescale for the EHV outflows
$t=R/V_{\rm ch}$.
When the map shows a clear separation between blue and red lobes,
the size of the outflow,
$R$, is calculated from the distance between red and blue peaks.
If the blue lobe is superposed on the red lobe,
we take the size of the FWHM contour as $R$.
Projection effects, not corrected for in our calculation,
can increase (if flow is in the plane of the sky) or decrease
(if flow is along the line of sight) the timescale by substantial factors.
In addition, $t$ would be smaller if we used the maximum velocity, as
is often done, rather than $V_{ch}$. On the other hand, Parker, Padman,
\& Scott (1991) have argued on the basis of detection statistics
that timescales derived from this method must underestimate
the true ages of outflows, by an order of magnitude on average.

\subsection {Other Information}

Since the discussion in \S\ 4  will need the luminosity and production rate
of ionizing photons, these are given in Table 6. The luminosity was
taken from the literature, corrected if necessary to the distances we
have adopted.
The rate of production of Lyman-continuum photons, $N_L$
was calculated from
\begin{equation}
N_L = 7.5\times 10^{43} \nu^{0.1} d^2 S_\nu T_e^{-0.45},
\end{equation}
where $N_L$ is in units of s$^{-1}$, $\nu$ the frequency in GHz,
$d$ the distance in kpc, $S_\nu$ the radio continuum flux in mJy,
and $T_e$ the electron temperature in units of 10$^4$ K (Rubin 1968).
We assumed $T_e$ = 8000 K and used the radio data given
in Table 6. To avoid optical depth problems, we used the highest frequency
radio continuum
data available which was not judged to be contaminated by dust emission.

In some of these sources, the radio continuum emission
may arise in an ionized stellar wind. In these cases, $N_L$ calculated
using equation (13)
will overestimate the production of Lyman-continuum photons
because Balmer-continuum photons may also ionize the dense winds
(Thompson 1984).  We have
computed the mass-loss rates in ionized winds (\mion ) by assuming that
all the radio
emission in Table~6 comes from an ionized wind.
We have used equation~(30) of
Panagia (1991), appropriate for a spherical, isothermal wind,
and assumed a terminal wind velocity of 300 {km~s$^{-1}$}\ and an
electron temperature of $10^4$ K. If the wind is confined to
a narrow jet, this equation will overestimate the mass-loss rate
(Reynolds 1986).  In addition, any contribution to $S_{\nu}$ from an optically
thin \hii\ region will cause us to overestimate \mion.

For comparison, we have also calculated the mass-loss rate in the
stellar wind based
on the HV CO emission, assuming momentum conservation, using
\begin{equation}
\dot M_{\rm CO} = {P \over t v_w},
\end{equation}
where $P$ and $t$ are the momentum and timescale for the HV flow, and
$v_w$ is the stellar wind velocity, assumed to be 300 {km~s$^{-1}$}.
\mion\ and $\dot M_{\rm CO}$ are given in Table~6.
In cases where our maps are incomplete, $\dot M_{\rm CO}$ will be
underestimated, but if Parker et al. (1991) are right about the ages
of outflows, $\dot M_{\rm CO}$ based on the usual method will be
overestimated. In defense of the traditional method, we note that
Natta \& Giovannardi (1990) derived mass loss rates from \ion{Na}{1} lines
for several stars which agreed very well with rates derived from CO.

\section {CONCLUSIONS}

In this section, we will summarize the results
and then consider different ideas for the origin of the EHV gas
and the nature of the stellar wind.

\subsection {Summary of Results}

Our detection of EHV wings was a result of serendipity
while conducting a CS study.
The selection criteria involved previous detections of dense gas
rather than any knowledge of molecular outflows,
although all the sources were known to have HV outflows.
Thus, it is probably significant
that almost all the sources in the study did show EHV wings.
EHV outflows are apparently common in regions similar to those in our sample,
dense regions of moderate to high mass star formation,
with luminosities ranging
from 500 {$L_\odot$}\ to 4{${} \times 10^{5}$}\ {$L_\odot$}.

The $J$ = 3{$\rightarrow$}2 line of CO shows the EHV wings
more clearly than lower $J$ lines,
indicating that it is a good way to search for EHV gas,
given an appropriate receiver and site.
The high ratios of the $J$ = 3{$\rightarrow$}2
to $J$ = 2{$\rightarrow$}1 EHV wing emission indicate
that the EHV gas is generally (GL 490 is an exception)
optically thin and warm ($T_{ex} >$ 20--90 K).
While the formal limits on $T_{ex}$ in Table 5 are not very high,
these limits assume the highest possible value
for the ratio of filling factors.
Lower values of this ratio would produce much higher $T_{ex}$,
comparable to the temperatures
needed to excite the CO \jj{14}{13}\ and \jj{16}{15}\ lines
observed toward many of these sources (Stacey 1992),
and impossibly high in some cases.
Howe et al. (1987) have detected \jj{14}{13}\ line of CO toward NGC~2071.
They remark that the warm component
responsible for the far-infrared CO emission
can account for the high-velocity gas seen in the millimeter CO lines
if the density is near 5{${} \times 10^{4}$}\ {cm$^{-3}$}\
and the temperature of the gas lies between 250 and 750 K.

The spectra indicate molecular gas moving at characteristic velocities
(in the sense of momentum over mass) of at least 20--40
{km~s$^{-1}$}\ with respect to the cloud rest velocity,
without projection corrections.
It is hard to accelerate molecular gas to such high velocities
without dissociation.
Based on the size of the outflows,
these velocities result in characteristic times of 1600--4200 yr.
Thus, these stars have been producing stellar winds,
probably neutral winds, within quite recent times,
even though many now are producing enough ionizing photons
to have substantial \hii\ regions.
One explanation, particularly attractive in distant, very obscured regions, is
that the source driving the outflow is separate
from the source producing the ionizing photons.
In the case of S140, however, the evidence points to IRS 1
as the source of the neutral wind,
even though it may be a main sequence star (Evans et al. 1989).

In contrast to recent spectroscopy
of nearby, less luminous sources with EHV gas,
our spectra are dominated by fairly smooth profiles,
with discrete velocity features being rare;
in addition the maps show clumpy emission only in the case of S140.
On the other hand, our maps are often incomplete or not completely sampled.
The weakness of the EHV wings, together with their likely high $T_{ex}$,
suggests a low filling factor, which could result from many small clumps.
In this case, the smooth profiles
could just be a result of inadequate spatial resolution.
Another shared spectral characteristic
is the presence of distinct changes of slope between the EHV and HV wings,
with the slope discontinuities in NGC 2071 and GL 2591
being the most dramatic.

\subsection {Nature of the EHV Gas}

We consider a quantitative test of the hypothesis
that the EHV gas traces the stellar wind.
Giovanardi et al. (1992) have shown that the
\hi\ winds in L1551 and HH 7--11 are sufficient to drive the HV molecular
outflow by momentum conservation. We apply a similar test to the
EHV gas, traced in our case by CO. In Table 6, we give $R_F$, the ratio
of the force available in
the EHV gas to that required to produce the momentum in the HV gas,
averaged over the lifetime of the HV flow. We also include $R_F$ for some
previously studied sources.

While the ratios for previously studied sources
are consistent with unity within likely uncertainties,
the ratios for our sources are not.
The largest ratio found in our observations is 0.10 (GL 490)
and the average is 0.045.
Low ratios appear to rule out the stellar wind interpretation for our sources,
but there are several reasons to suspend judgement.
Since the EHV gas is spread over a large range of velocities,
sensitivity limitations could cause an underestimate of the emission.
Second, our division between HV and EHV gas is not precise
in sources without clear spectral breaks;
moving the boundary between HV and EHV gas to lower velocities
would raise the ratios.
Examples of this effect can be seen in the different ratios
for \hbox{HH 7--11} and IRAS~03282+3035,
depending on whether the intermediate-velocity wing
is assigned to the HV or EHV wings.
Third, by drawing our boundaries in velocity
rather than fitting a series of Gaussians,
we are neglecting gas at low {\it projected\/} velocities,
a more serious omission for the EHV gas than for the HV gas.
Fourth, incomplete maps in some sources introduce uncertainties;
for example, the data of Chernin \& Masson (1992)
yield $R_F$ about 3 times our value for NGC 2071.
These four factors typically introduce factor of 2--3 uncertainties,
so are unlikely to change the conclusions.

Perhaps more important is our assumption
that all the C is associated into CO in calculating $R_F$
for the stellar wind hypothesis.
If, on average, only 4.5\% of the C is in CO,
the average $R_F$ would be unity.
The chemical calculations of Glassgold et~al. (1991)
find CO abundances as low as [CO]/[C] = 0.04
only for low mass-loss rates
(\mdot\ {$\rlap{\raise.4ex\hbox{$<$}}\lower.55ex\hbox{$\sim$}\,$}\
3{${} \times 10^{{-6}}$}\ \msunyr),
but our EHV outflow sources have much larger mass-loss rates
(\mdot\ $>$ 10$^{-4}$ \msunyr).
However, the calculations of Glassgold et~al. (1991)
were for cool ($T_{\star}$ = 5000~K) stars,
while the stars in this study are much more luminous
and most are producing some ionizing photons.
Ionizing and dissociating photons from these stars
may cause lower CO abundances in the wind.
Glassgold et al. (1991) did consider a model
with \mdot\ = 3{${} \times 10^{{-6}}$}\ \msunyr\
which had a \hbox{$T_\star$ = 6000~K}.
Even this modest increase in $T_\star$
caused the CO abundance to drop by three orders of magnitude.
Glassgold et al. (1991) also note
that they have ignored dust formation in the wind
which could shield molecules.

We need models of winds from stars with $L > 500$ {$L_\odot$}\
and higher temperatures to determine CO abundances in the winds.
Empirically, we can assess
whether the nature of the stars may be affecting the CO abundance
by plotting $R_F$ versus luminosity and versus $N_L$,
the rate of production of ionizing photons (Fig. 9).
The plots indicate that $R_F$ declines sharply with increasing luminosity
and increasing $N_L$, with the latter trend being particularly striking.
If the CO abundance decreases with ionizing photon rate,
the EHV CO emission could still be tracing the stellar wind.
W28 A2 is unusual in having an impressive EHV wing
while producing a large $N_L$.
As a distant and complex region,
W28 A2 may be a case in which the outflow is driven by a separate source.
However, the outflow must still avoid ionization
by the source of the ionizing photons for the extended \hii\ region.

We conclude that an origin for the EHV wings in molecules formed
in a neutral wind cannot be clearly ruled out. Even if the EHV CO does
arise in the wind, it is not a useful probe of the momentum in high
luminosity sources, since the CO is unlikely to be fully associated.

An alternative origin for the EHV gas is ambient gas
which has been entrained or swept-up by the wind.
In cases where there is a break in the slope in the profile,
the process must be distinct from that which produces the HV wing.
Chernin \& Masson (1992) argue
that it is difficult to accelerate molecules
to the extremely high velocities without dissociation,
so they favor a picture in which the EHV molecules
have reformed after the passage of a fast shock.
Models of such shocks indicate that the molecules reform
when the column density of the shocked material
reaches about 1--3{${} \times 10^{{20}}$}\ \cmc\
(Neufeld \& Dalgarno 1989a; Hollenbach \& McKee 1989).
By $N$ = 3{${} \times 10^{{20}}$}\ \cmc, the carbon is fully associated,
with an abundance of 4{${} \times 10^{{-4}}$},
implying a CO column density of about 1{${} \times 10^{{17}}$}\ \cmc,
and the gas has a temperature of about 500 K.
The high temperature is consistent with observations of high-$J$ CO emission
in several of our sources (Howe et~al. 1987; Stacey et al. 1987)
and with our high \rco,
but our column density estimates for the EHV CO at the central positions
range from 2{${} \times 10^{{15}}$}\ to 1.2{${} \times 10^{{17}}$}\ \cmc,
with only W28~A2 having the column density expected in this picture.
Larger column densities may be present
if the emission comes from many small regions,
as seems likely on other grounds.
The recent detection of SO in NGC 2071 clearly indicates
that shock chemistry is active,
but the spatial and velocity structure of the SO emission
is hard to interpret simply in terms of shock models
(Chernin \& Masson 1993).

In the picture of outflows proposed by Stahler (1993), the stellar wind
sweeps up and entrains gas in a turbulent mixing process, which may allow
acceleration to EHV velocities without dissociation. In this case, the
original molecular cloud abundances may be preserved, with only 12\%
of the C in CO. We can then compare the mass, momenta, and energy
of the EHV gas to that of the HV gas, using the second columns for
each in Table 5. It is clear that the HV gas carries most of the mass
and momentum in all cases; in a few cases, the energy in the EHV wings
is comparable to that in the HV wings, but it is usually not. Thus,
the EHV wings, if swept-up matter, do not change the overall energetics
of the outflows. The $R_F$ values are still less than 1 (Table 6),
but the ratios are now sufficiently close (0.36 in the average) that
uncertainties could explain the difference. In the picture of Stahler,
in which momentum is transferred through turbulent mixing,
one might expect a similar force to have been applied to gas at
all velocities.

A different model of outflows, with collimated jets as the driving source,
has been suggested recently (Masson \& Chernin 1993; Raga et al. 1993).
Jets are plausible driving sources only if Parker et al. (1991) are
correct in arguing that traditional analyses have  underestimated
lifetimes and overestimated $\dot M_{\rm CO}$ by a factor of 10.
In the jet models, the EHV CO emission is produced by the bow shock,
where the jet encounters the ambient cloud. This simple model would
predict a single EHV feature at the end of each lobe, rather than
an extended wing, but Masson \& Chernin (1993) proposed that the jets wander,
producing multiple EHV blobs. Whether this model can reproduce the
rather smooth EHV wings and the large velocity extents that we observe
remains to be seen.

In general, the picture of the EHV wings as gas originally in the ambient cloud
is certainly plausible.
However, this picture provides no obvious explanation
for the rather striking trend of $R_F$ with $N_L$ in Figure~9,
nor for the fact that $R_F \sim 1$
in some sources with low luminosity and $N_L$.

\subsection {Where is the Carbon in the Wind?}

If the EHV material is, in fact, ambient material
swept up by a faster stellar wind,
we would still need to find the underlying stellar wind
driving both the HV and EHV components.
In young low-luminosity stars,
ionized winds, if they are present at all,
are insufficient to sweep up the molecular gas (Evans et~al. 1987).
The situation for more luminous objects is less clear.
In Table~6, we compare
mass-loss rates inferred from the radio emission (\mion )
with the stellar wind $\dot M$ inferred from the HV CO ($\dot M_{\rm CO}$)
assuming momentum conservation (eq. [14]).
Only GL 490 and radio component 2 in Cepheus A
actually have the properties expected for spherical winds,
and \mion\ for those sources
is an order of magnitude less than $\dot M_{\rm CO}$.
The estimates of \mion\ are upper limits for the other sources,
and $\dot M_{\rm CO}$ estimates are often lower limits.
Still, only W28 A2 and GL 2591 have  \mion\ comparable to $\dot M_{\rm CO}$.
The radio source in GL~2591 does not coincide with the infrared source,
making its status unclear.
W28~A2 has been classified as an ultracompact \hii\ region
with a shell structure (Wood \& Churchwell 1989)
and its radio emission is too strong to arise in an ionized stellar wind.

The best interpretation of the radio continuum results for our sources
is that the true value of \mion\ is much less than
$\dot M_{\rm CO}$ in all of the sources in the sample.  This leads to
the conclusion that the winds are primarily neutral, even
in these more luminous objects. Detection of \hi\ wings toward
DR 21 (Russell et al. 1992) supports the idea that luminous stars
may also have neutral winds. How a star producing a substantial
number of ionizing photons also drives a neutral stellar wind is a puzzle
worthy of theoretical consideration. We will confine ourselves to asking
where is the carbon in the stellar wind? Since these are oxygen-rich objects,
the carbon is unlikely to be found in solid particles, leaving CO, C$^0$,
and C$^+$ as possibilities.

In low luminosity sources, the carbon may well be in CO
and already detected since $R_F \sim 1$.
Observations of \hi\ in some low-luminosity sources
over large velocity intervals indicates
that there is atomic or molecular material
at velocities up to \hbox{150 {km~s$^{-1}$}} from line center
(Lizano et~al. 1988; Giovanardi et al. 1992).
Could we have detected wings of this width
if the gas contains enough CO to make $R_F = 1$?
A wind with a rectangular profile
and a terminal velocity of 150 {km~s$^{-1}$}\
would produce a brightness temperature less than 0.2 K
in the $J$ = 3{$\rightarrow$}2 line of CO for all sources except W28 A2,
where it would be more like 2~K.
Given that $\pm$150 {km~s$^{-1}$}\ would extend to the ends of our baseline
and given the systematic uncertainties in the baselines,
it would be difficult for us to rule out
the existence of this kind of CO component,
even if the spectral shape had some degree of central peaking.
Continuing improvements in instrumentation and observing techniques
may eventually permit a meaningful search
for even higher velocity molecular gas.

Another possibility is that the C is neutral and atomic.
We have estimated the strength of the EHV
wings in $^3P_1${$\rightarrow$}$^3P_0$ for the
sources in our sample, assuming that the gas is hotter than 50 K,
all of the carbon not in CO is in C$^0$, and the force in the EHV wind
matches that in the molecular HV wind.  With these assumptions, it should
be possible to detect C$^0$ emission from the EHV gas
in at least half the sources in our sample.
For most, the [\ion{C}{1}] line would be detectable
at some EHV velocities even if [C$^0$]/[C] is as low as 0.1.
Walker et~al. (1993) have observed HV wings
in the $^3P_1${$\rightarrow$}$^3P_0$ transition of \ion{C}{1} at 492 GHz
in several sources,
including three of the sources discussed in this paper.
They found that the C$^0$ abundance
was comparable to the abundance in quiescent cloud material.
Walker et al. (1993) would  not have been able to detect EHV wings,
but longer integration time under good conditions should be able
to determine if a substantial fraction of the carbon
in the stellar wind is C$^0$.

The carbon could also be in the form of C$^+$, if it is ionized by
stellar photons.  High-velocity shocks could also produce substantial
ultraviolet emission, but shock models for dense gas predict C$^+$ column
densities which are always less than those of CO (Neufeld \& Dalgarno 1989b).
The strong trend for $R_F$
to decrease as $N_L$ increases may indicate
a chemical shift from CO to C$^+$, due to the stellar ultraviolet emission.
Estimates based on assumptions similar to those used for [\ion{C}{1}]
predict strong [\ion{C}{2}] emission from a mostly
C$^+$ wind.  The sensitivity of current airborne spectrometers
to the 158 {$\mu$m}\ $^2P_{3/2}${$\rightarrow$}$^2P_{1/2}$ line
is sufficient, in principle,
to place meaningful limits on the amount of C$^+$ in the EHV winds in
our sources, but practical limitations would make it very hard
to detect very weak, broad wings on a narrow bright line in the far-infrared.

While the observations are difficult, failure to detect C in any
form at the level required by the usual analysis of the CO HV flows
would support the idea that outflow ages have been underestimated
and lend credence to models of jet-driven outflows.

\acknowledgements

We thank John E. Howe, Jeffrey G. Mangum, Ren\'e Plume,
Constance E. Walker, Yangsheng Wang, Lianzhou Yu, and Shudong Zhou
for acquiring some of the data.
We wish to thank L. M. Chernin for helpful comments.
This work was supported by NSF grant AST-9017710,
by a grant from the W. M. Keck Foundation,
by a grant from the Texas Advanced Research Program
to the University of Texas,
and by a David and Lucile Packard Foundation Fellowship.

\appendix

\section{The Behavior of $^{12}R_{32}$}

To understand the general behavior of \rco,
we can simplify equation (2) by assuming
that the optical depths follow the LTE relation (eq. [4]).
The largest possible value of the ratio of beam filling factors, $\phi$
is (3/2)$^2$, assuming a point source centered in both beams.
The resulting \rco\ is plotted in Figure~10
as a function of the optical depth
for various values of excitation temperature.
For an optically thick cloud, \rco\ is quite insensitive
to both $\tau$ and $T_{ex}$.
For an optically thin cloud, \rco\ can be
as low as 0.44 (for $T_{ex} = 10$ K and $\phi$ = 1)
and as high as (3/2)$^4$ [for \hbox{$T_{ex}$ {$\rightarrow$}\ $\infty$}
and $\phi$ = (3/2)$^2$] depending on the temperature and source extent.
Therefore, \rco\ larger than $\sim$$(3/2)^2$ implies
that the cloud is optically thin,
but \rco\ smaller than $\sim$$(3/2)^2$ does not give any clue
to the physical conditions unless the source extent is known.

One might regard \rco\ as a good temperature indicator
for optically thin clouds based on Figure 10,
but it is hard to determine $\phi$, the ratio of beam filling factors.
The lower limits to $T_{ex}$ in Table 5 were obtained
by assuming a point source. In addition, mixtures of gas at different
temperatures will result in an estimate of $T_{ex}$ less than that of
the hotter gas. If both components have equal optical depth, the
derived temperature is close to the average of the two components, according
to simulations that we have done. Simulations with two equal temperature
components of differing optical depth, but both still optically thin,
show that the true temperature will be somewhat underestimated by this
method.  Thus, the excitation temperatures we have derived from this method
should be regarded as quite strong lower limits.

\section{Temperature Estimates from the CO Optical Depth Ratio}

It is in principle possible to determine the excitation temperature
from the ratio of the optical depths of two adjacent transitions assuming LTE
and, in the process, avoid some of the problems with beam filling factors
alluded to in Appendix A.
Consider two transitions, CO $J$+1 {$\rightarrow$}\ $J$ (labeled by $u$)
and CO \hbox{$J$ {$\rightarrow$}\ $J$--1} (labeled by $l$); then
\begin{equation}
{{\tau _u}\over {\tau _l}} = {{J+1}\over J}\
   {{1-\exp\left(-{{h\nu_u}/{kT_{\rm ex}}}\right)}
   \over {\exp\left({h\nu_l}/{kT_{\rm ex}}\right)-1}}.
\end{equation}
Snell et al. (1984) and Margulis \& Lada (1985), using this method
for several outflows with CO \jj21\ and CO \jj10\ data
(calculating $\tau_{21}$ and $\tau_{10}$ from $^{iso}R_2$ and $^{iso}R_1$
using eq.~[1]),
found oddly low excitation temperatures ($T_{ex}$ = 6.5--25 K)
except in Ori~A.
Although we have determined optical depths of
CO $J$ = 3{$\rightarrow$}2 and CO $J$ = 2{$\rightarrow$}1 lines
for most of the inner wings and some of the outer wings,
we did not use this method to calculate the excitation temperatures.
The reason is that outflows may be clumpy
and there may be several unresolved components in our main beam.

Assuming that there are two components, A and B, in our main beam,
one with $\tau _{u{\rm A}}$ and the other with $\tau _{u{\rm B}}$,
but both with the same excitation temperature, $T_{ex}$,
then $\tau _{l{\rm A}}$ and $\tau _{l{\rm B}}$ can be
deduced by equation~(B1).
The observed radiation temperature is
roughly the sum of those from each component.
The procedure would derive the optical depths,
$\tau _{u,obs}$ and $\tau _{l,obs}$,
assuming all the filling factors are about the same, from
\begin{equation}
{{1-\exp(-\tau _{\rm obs})}\over {1-\exp(-\tau _{\rm obs}/X)}}
   = {{1-\exp(-\tau _{\rm A})\ +\ 1-\exp(-\tau _{\rm B})}\over
     {1-\exp(-\tau _{\rm A}/X)\ +\ 1-\exp(-\tau _{\rm B}/X)}},
\end{equation}
where $\tau$ can be either $\tau _u$ or $\tau _l$.

The resulting $T_{ex,obs}$ (excitation temperature calculated by eq. [B1])
is shown as a function of $T_{ex}$
(real excitation temperature of the cloud) in Figure~11, assuming two
components with equal filling factor
with one component having \hbox{$\tau$ = 1}
and the other \hbox{$\tau$ = 10}.
These curves clearly show that this method can fail badly
as a temperature estimator if there are two or more components
of different optical depth in the beam.  The qualitative result is true
for any pair of opacities as long as they differ by a factor of 2--3 or
more and the component with larger opacity has $\tau>1$.
If the low temperatures derived by this method
from the \hbox{$J$ = 2{$\rightarrow$}1} and \hbox{$J$ = 1{$\rightarrow$}0} data
are an artifact of using the technique
in the presence of inhomogeneities in the sources,
Figure~11 shows that a similar analysis
using the \hbox{$J$ = 3{$\rightarrow$}2}
and \hbox{$J$ = 2{$\rightarrow$}1} transitions
should yield higher temperatures.  Indeed, the typical $T_{ex}$'s
we derive by this method are twice those derived in the earlier work
with the \hbox{$J$ = 2{$\rightarrow$}1}
and \hbox{$J$ = 1{$\rightarrow$}0} lines.

\clearpage
\setbox1=\vbox{\tablesize\halign{
\hfil#\hfil&\quad\hfil#&#\hfil&\quad\hfil#\hfil
&\quad\hfil#\hfil&\quad\hfil#&#\hfil&\quad\hfil#\hfil\cr
\multispan8 \hfil TABLE 1 \hfil \cr
\cr
\multispan8 \hfil \tup{O}{BSERVATIONAL}\ \tup{P}{ARAMETERS} \hfil \cr
\cr
\noalign{\hrule \vskip 2pt \hrule \vskip 2pt}
Telescope&\multispan2 \hfil Transition \hfil
&$\nu$&$\Delta v^a$&\multispan2 \hfil $\theta^b$ \hfil &$\eta _{mb}^c$\cr
&&&(GHz)&({km~s$^{-1}$})\cr
\noalign{\vskip 2pt \hrule \vskip 2pt}
CSO&$^{13}$&CO $J$ = 2{$\rightarrow$}1&220.3987&1.37&32&\arcsec &0.71\cr
   &       &CO $J$ = 2{$\rightarrow$}1&230.5380&1.28&30&        &0.71\cr
   &$^{13}$&CO $J$ = 3{$\rightarrow$}2&330.5880&0.91&21&        &0.55\cr
   &       &CO $J$ = 3{$\rightarrow$}2&345.7960&0.86&20&        &0.55\cr
\noalign{\vskip 2pt \hrule \vskip 10pt}}}
\setbox2=\vbox{\tablesize\hsize=\wd1\goodfil
$^a$ Velocity resolution in {km~s$^{-1}$}. \par
$^b$ Beam size (FWHM) in arcseconds. \par
$^c$ Main beam efficiency (Mangum 1993).}
\vbox{{\box1}{\box2}}
\vskip 2cm
\setbox1=\vbox{\tablesize\halign{
#\hfil&\quad#\hfil&#\hfil&#\hfil
&\hfil\quad#&#\hfil&#\hfil&#\hfil&\quad#\hfil&\hfil#
&\quad\hfil#&#\hfil&\quad\hfil#&#\hfil&\quad#\hfil\cr
\multispan{15} \hfil TABLE 2 \hfil \cr
\cr
\multispan{15} \hfil \tup{S}{OURCE}\ \tup{P}{ARAMETERS} \hfil \cr
\cr
\noalign{\hrule \vskip 2pt \hrule \vskip 2pt}
Source&\multispan3 \hfil $\alpha _{1950}$\hfil
&\multispan4 \hfil $\delta _{1950}$\hfil
&$d^a$&References&\multispan2 \hfil $\Delta \alpha ^b$\hfil
&\multispan2 \hfil $\Delta \delta ^b$\hfil &Line$^c$\cr
&&&&&&&&(kpc)\cr
\noalign{\vskip 2pt \hrule \vskip 2pt}
W3 IRS 5 &02$^h$&21$^m$&53$^s\!\!$.1
         &61&$^{\circ }$&52${'}$&22${''}$&2.3
         &1&2400&${''}$&0&${''}$
         &3--2, $^{13}$3--2, $^{13}$2--1\cr
         &&&&&&&&&&  1200& &    0& &2--1\cr
GL 490   &03&23&39.2 &  58&&36&36 &1.0  &2&--1800& &    0&
         &3--2, 2--1, $^{13}$3--2, $^{13}$2--1\cr
NGC 2071 &05&44&30.6 &  00&&20&42 &0.39 &3&--2700& &  900&
         &3--2, $^{13}$3--2, $^{13}$2--1\cr
         &&&&&&&&&&--2700& &    0& &2--1\cr
W28 A2   &17&57&26.8 &--24&&03&54 &3.0  &4& --600& &    0&
         &3--2, 2--1, $^{13}$3--2, $^{13}$2--1\cr
GL 2591  &20&27&35.8 &  40&&01&14 &2.0  &5&     0& &10800&
         &3--2, 2--1, $^{13}$3--2, $^{13}$2--1\cr
S140     &22&17&41.2 &  63&&03&45 &0.91 &6& --600& &--600&
         &3--2, 2--1, $^{13}$3--2, $^{13}$2--1\cr
Cepheus A&22&54&19.0 &  61&&45&47 &0.725&7& --780& &    0&
         &3--2, 2--1, $^{13}$3--2\cr
         &&&&&&&&&&     0& & 1200& &$^{13}$2--1\cr
\noalign{\vskip 2pt \hrule \vskip 10pt}}}
\setbox2=\vbox{\tablesize\hsize=\wd1\goodfil
$^a$ Distance from the Sun in kpc. \par
$^b$ Coordinates of {\tabsmall OFF} positions
     relative to the central position. \par
$^c$ 3--2: CO \jj32; \quad $^{13}$3--2: $^{13}$CO \jj32; \quad
     2--1: CO \jj21; \quad $^{13}$2--1: $^{13}$CO \jj21. \par
\refs (1) Georgelin \& Georgelin 1976; (2) Snell et al. 1984;
(3) Anthony-Twarog 1982; (4)~Harvey \& Forveille 1988;
(5) Merrill \& Soifer 1974; (6) Crampton \& Fisher 1974;
(7)~Blaauw, Hiltner, \& Johnson 1959.}
\vbox{{\box1}{\box2}}
\vfill\eject

\setbox1=\vbox{\tablesize\halign{
#\hfil&\quad\hfil#&\quad#\hfil&\quad\hfil#
&\quad\hfil#\hfil&\quad\hfil#\hfil\cr
\multispan6 \hfil TABLE 3 \hfil \cr
\cr
\multispan6 \hfil \tup{L}{INE}\ \tup{P}{ARAMETERS} \hfil \cr
\cr
\noalign{\hrule \vskip 2pt \hrule \vskip 2pt}
Source&Line$^a$&\multispan2 \hfil Baseline$^b$ \hfil
&Noise$^c$(K)&Threshold$^d$(K)\cr
\noalign{\vskip 2pt \hrule \vskip 2pt}
W3 IRS 5 &3--2        &(--150, --90)&(10, 40)&0.115&0.264\cr
         &2--1        &(--150, --90)&(30, 100)&0.038&0.096\cr
         &$^{13}$3--2 &(--150, --70)&(0, 100)&0.116&0.285\cr
         &$^{13}$2--1 &(--150, --70)&(0, 30)&0.019&0.048\cr
\tabvs
GL 490   &3--2        &(--100, --75)&(50, 100)&0.056&0.126\cr
         &2--1        &(--100, --80)&(50, 100)&0.155&0.297\cr
         &$^{13}$3--2 &(--100, --50)&(40, 100)&0.291&0.645\cr
         &$^{13}$2--1 &(--100, --50)&(40, 100)&0.114&0.264\cr
\tabvs
NGC 2071 &3--2        &(--130, --60)&(100, 150)&0.091&0.195\cr
         &2--1        &(--130, --60)&(90, 150)&0.035&0.084\cr
         &$^{13}$3--2 &(--130, --60)&(90, 150)&0.064&0.135\cr
         &$^{13}$2--1 &(--130, --50)&(60, 150)&0.035&0.084\cr
\tabvs
W28 A2   &3--2        &(--100, --90)&(180, 220)&0.225&0.579\cr
         &2--1        &(--200, --100)&(100, 150)&0.090&0.228\cr
         &$^{13}$3--2 &(--180, --80)&(70, 180)&0.222&0.546\cr
         &$^{13}$2--1 &(--150, --100)&(100, 150)&0.211&0.516\cr
\tabvs
GL 2591  &3--2        &(--120, --50)&(25, 85)&0.211&0.528\cr
         &2--1        &(--120, --50)&(25, 85)&0.070&0.180\cr
         &$^{13}$3--2 &(--120, --50)&(25, 85)&0.189&0.432\cr
         &$^{13}$2--1 &(--120, --40)&(20, 85)&0.026&0.072\cr
\tabvs
S140     &3--2        &(--90, --60)&(30, 80)&0.084&0.219\cr
         &2--1        &(--100, --60)&(30, 100)&0.046&0.117\cr
         &$^{13}$3--2 &(--100, --60)&(30, 100)&0.078&0.180\cr
         &$^{13}$2--1 &(--100, --60)&(30, 100)&0.037&0.102\cr
\tabvs
Cepheus A&3--2        &(--200, --150)&(80, 120)&0.422&0.900\cr
         &2--1        &(--150, --100)&(100, 150)&0.285&0.639\cr
         &$^{13}$3--2 &(--100, --50)&(30, 100)&0.120&0.306\cr
         &$^{13}$2--1 &(--100, --50)&(30, 100)&0.027&0.069\cr
\noalign{\vskip 2pt \hrule \vskip 10pt}}}
\setbox2=\vbox{\tablesize\hsize=\wd1\goodfil
$^a$ See Table 2. \par
$^b$ Intervals considered in the baseline determination in {km~s$^{-1}$}. \par
$^c$ RMS of noise with the resolution in Table 1. \par
$^d$ Thresholds for the line ratios in data with $\Delta v$ = 2 {km~s$^{-1}$}.}
\vbox{{\box1}{\box2}}
\vfill\eject

\setbox1=\vbox{\tablesize\halign{
#\hfil&\quad\hfil#&\hfil#\hfil&\hfil#&\hfil#\hfil&\hfil#
&\quad\hfil#
&\quad\hfil#&\hfil#\hfil&\hfil#&\hfil#\hfil&\hfil#
&\quad\quad\hfil#&#\cr
\multispan{14} \hfil TABLE 4 \hfil \cr
\cr
\multispan{14} \hfil \tup{W}{ING}\ \tup{B}{OUNDARIES} \hfil \cr
\cr
\noalign{\hrule \vskip 2pt \hrule \vskip 2pt}
\tup{S}{OURCE}&\multispan2 &\hfil \tup{B}{LUE}\hfil&\multispan2
&&\multispan2 &\hfil \tup{R}{ED}\hfil&\multispan2
&\multispan2 \hfil \tup{F}{ULL}\ \tup{W}{IDTH}\hfil\cr
&&Outer&&Inner&&\hfil $V_0$\hfil
&&Inner&&Outer&&\multispan2 \hfil \tup{}{OF}\ \tup{W}{ING}\hfil\cr
\noalign{\vskip 2pt \hrule \vskip 2pt}
W3 IRS 5     &--76&&--60$^a$&&--50&--37&--28&&--18$^a$&&--4& 72\cr
GL 490$^d$   &--76&&--40$^b$&&--16&--13& --8&&  14$^b$&& 36&112\cr
NGC 2071$^e$ &--50&&--10$^a$&&   0&  11&  18&&  28$^a$&& 68&118\cr
W28 A2$^f$   &--70&&--20$^c$&&   2&   9&  18&&  32$^c$&& 70&140\cr
GL 2591$^g$  &--49&&--21$^a$&&--11& --6&   3&&  15$^a$&& 25& 74\cr
S140$^h$     &--56&&--30$^b$&&--12& --7& --4&&   6$^a$&& 30& 86\cr
Cepheus A$^i$&--81&&--29$^a$&&--17&--12& --5&&   7$^a$&& 51&132\cr
\noalign{\vskip 2pt \hrule \vskip 10pt}}}
\setbox2=\vbox{\tablesize\hsize=\wd1\goodfil
\note All velocities are in {km~s$^{-1}$}.

$^a$ The boundaries between inner and outer wings were set
by noticing distinct changes of the slope in the CO \jj32\ spectra.

$^b$ The CO \jj32\ spectra do not show clear change of slope,
and the inner wings were defined from previous observations of HV outflows.

$^c$ The CO \jj32\ spectra do not show clear change of slope,
and the boundaries between inner and outer wings were set subjectively.

$^d$ Our inner wings coincide with the wings seen by Snell et~al. (1984).
The full widths of the EHV CO \jj10\ line
of Koo (1989) and Margulis \& Snell (1989) are
wider than ours by 40 and 13 {km~s$^{-1}$}, respectively,
probably because of higher signal-to-noise ratios.

$^e$ Our inner wings coincide with
the entire wings seen by Bally (1982),
while our outer wings are wider by \hbox{$\sim$40 {km~s$^{-1}$}} on each side,
similar to those of Chernin \& Masson (1992).

$^f$ The full width of the wings is
comparable to the width in CO \jj10\ line (Harvey \& Forveille 1988).

$^g$ Our inner wings have the same boundaries
as defined in Lada et~al. (1984)
and are similar to those of Mitchell, Maillard, \& Hasegawa (1991).

$^h$ Our inner blue wing has the same extent as the blue wing
of Hayashi et al. (1987) and Snell et al. (1984).
The full width of the EHV CO \jj10\ line of Koo (1989) is
wider than ours by 30 {km~s$^{-1}$},
but that of Margulis \& Snell (1989) is narrower than ours by 11 {km~s$^{-1}$}.

$^i$ Our inner wings correspond to the wings shown
in \hbox{Figs. 3{\it b\/}--3{\it g\/}} of Ho et al. (1982).
Our outer blue wing is wider than the wing
in Fig. 3{\it h\/} of Ho et al. (1982),
and our outer red wing was not seen by them.}
\vbox{{\box1}{\box2}}
\vfill\eject

\setbox1=\vbox{\tablesize\halign{
#\hfil&\quad#\hfil&\quad\hfil#&\quad\hfil#
&\quad\hfil#&#\hfil&\quad\hfil#&#\hfil&\quad\hfil#&#\hfil
&\quad\hfil#&#\hfil&\quad\hfil#&#\hfil&\quad\hfil#&#\hfil\cr
\multispan{16} \hfil TABLE 5 \hfil \cr
\cr
\multispan{16} \hfil \tup{P}{HYSICAL}\ \tup{P}{ARAMETERS}
                     \tup{}{OF}\ \tup{O}{UTFLOWS} \hfil \cr
\cr
\noalign{\hrule \vskip 2pt \hrule \vskip 2pt}
&&&\hfil $T_{ex}^b$\hfil
&\multispan2 \hfil $M^c$\hfil &\multispan2 \hfil $M^d$\hfil
&\multispan2 \hfil $P^c$\hfil &\multispan2 \hfil $P^d$\hfil
&\multispan2 \hfil $E_K^c$\hfil &\multispan2 \hfil $E_K^d$\hfil\cr
Source &Wing &$X^a$&\hfil (K)\hfil
&\multispan2 \hfil($M_\odot$)\hfil
&\multispan2 \hfil($M_\odot$)\hfil
&\multispan4 \hfil($M_\odot$ {km~s$^{-1}$})\hfil
&\multispan4 \hfil(10$^{45}$ erg)\hfil\cr
\noalign{\vskip 2pt \hrule \vskip 2pt}
W3 IRS 5&Outer blue&66&\na &0&.014&0&.11&0&.38& 3&.1&0&.11& 0&.87\cr
        &Inner blue&  &\na & &\na &1&.5 & &\na&24&. & &\na& 4&.0 \cr
        &Inner red &  &\na & &\na &6&.2 & &\na&78&. & &\na&10&.  \cr
        &Outer red &  &\na &0&.026&0&.21&0&.63& 5&.0&0&.15& 1&.2 \cr
\tabvs
GL 490$^e$&Outer blue&61&\na &0&.0084&0&.067&0&.32& 2&.5&0&.13&1&.0 \cr
          &Inner blue&  &\na & &\na  &2&.3  & &\na&19&. & &\na&2&.1 \cr
          &Inner red &  &\na & &\na  &1&.9  & &\na&18&. & &\na&2&.1 \cr
          &Outer red &  &\na &0&.018 &0&.14 &0&.60& 4&.8&0&.21&1&.6 \cr
\tabvs
NGC 2071$^e$&Outer blue&73&$>$17&0&.0049&0&.039&0&.14&1&.1 &0&.043&0&.34\cr
            &Inner blue&  & \na & &\na  &0&.28 & &\na&3&.9 & &\na &0&.55\cr
            &Inner red &  & \na & &\na  &0&.28 & &\na&2&.7 & &\na &0&.28\cr
            &Outer red &  & \na&0&.0029&0&.023&0&.083&0&.67&0&.027&0&.21\cr
\tabvs
W28 A2&Outer blue&40&$>$94&0&.16& 1&.3&7&.0 & 56&.&3&.2 &26&.\cr
      &Inner blue&  & \na & &\na&38&. & &\na&500&.& &\na&75&.\cr
      &Outer red &  & \na &0&.14& 1&.1&4&.6 & 37&.&1&.5 &12&.\cr
\tabvs
GL 2591$^e$&Outer blue&55&$>$16&0&.030 & 0&.24 &0&.66&  5&.3 &0&.15 & 1&.2 \cr
           &Inner blue&  & \na & &\na  &31&.   & &\na&240&.  & & \na&19&.  \cr
           &Inner red &  & \na & &\na  & 2&.1  & &\na& 23&.  & & \na& 2&.7 \cr
           &Outer red &  & \na &0&.0048& 0&.038&0&.12&  0&.94&0&.029& 0&.23\cr
\tabvs
S140&Outer blue&58&$>$23&0&.0072& 0&.057&0&.23& 1&.9&0&.080&0&.64\cr
    &Inner blue&  & \na & &\na  &12&.   & &\na&93&. & &\na &7&.6 \cr
    &Inner red &  & \na & &\na  &13&.   & &\na&64&. & &\na &3&.5 \cr
    &Outer red &  & \na &0&.022 & 0&.18 &0&.42& 3&.3&0&.085&0&.68\cr
\tabvs
Cepheus A$^e$&Outer blue&58& \na &0&.0045&0&.036&0&.14& 1&.2&0&.053&0&.42\cr
             &Inner blue&  & \na & &\na  &1&.6  & &\na&13&. & &\na &1&.1 \cr
             &Inner red &  & \na & &\na  &6&.9  & &\na&65&. & &\na &6&.4 \cr
             &Outer red &  &$>$17&0&.027 &0&.22 &0&.81& 6&.5&0&.27 &2&.1 \cr
\noalign{\vskip 2pt \hrule \vskip 10pt}}}
\setbox2=\vbox{\tablesize\hsize=\wd1\goodfil
$^a$ The isotopic abundance ratio, [CO]/[$^{13}$CO]
     (Langer \& Penzias 1990). \par
$^b$ Lower limits of $T_{ex}$ calculated
     from eq. (3) assuming $\phi = (3/2)^2$. \par
$^c$ Assuming that the sources are neutral atomic wind. \par
$^d$ Assuming that the sources are swept-up molecular gas. \par
$^e$ Maps cover only the central portions of the outflows. \par}
\vbox{{\box1}{\box2}}
\vfill\eject

{\newdimen\rotdimen
\def\vspec#1{\special{ps:#1}}
\def\rotstart#1{\vspec{gsave currentpoint currentpoint translate
   #1 neg exch neg exch translate}}
\def\rotfinish{\vspec{currentpoint grestore moveto}}
\def\rotl#1{\rotdimen=\ht#1\advance\rotdimen by\dp#1%
   \hbox to\rotdimen{\vbox to\wd#1{\vskip\wd#1\rotstart{270 rotate}%
   \box#1\vss}\hss}\rotfinish}%
\def\tabnormal{\scriptsize}
\def\tabsmall{\tiny}
\def\tabbaseskip{\baselineskip=11pt}
\setbox1=\vbox{\tablesize
\def\hfin{\hskip 43pt plus 20pt minus 20pt}
\halign{
#\hfil&\quad\hfil#&#\hfil&\quad\hfil#&#\hfil&\quad\hfil#&\quad
\hfil#&\quad\hfil#&\quad\hfil#&#&\quad\hfil#&\quad\hfil#&\quad
\hfil#&#&\quad\hfil#&\quad\hfil#&\quad\hfin#&#\hfil\cr
\multispan{18} \hfil TABLE 6 \hfil \cr
\cr
\multispan{18} \hfil \tup{R}{ATIOS}\ \tup{}{OF}
                     \tup{D}{RIVING}\ \tup{F}{ORCES}\ \tup{}{AND}
                     \tup{R}{ELATED}\ \tup{Q}{UANTATIES} \hfil \cr
\cr
\noalign{\hrule \vskip 2pt \hrule \vskip 2pt}
&\multispan6\quad\hfil EHV CO \hfil&\hfil HV CO \hfil
&\multispan3\quad\hfil Far-Infrared \hfil
&\multispan7\quad\hfil Radio Continuum\hfil\cr
\noalign{\vskip -9pt}
&\multispan6\quad\hrulefill&\hrulefill
&\multispan3\quad\hrulefill&\multispan7\quad\hrulefill\cr
\tup{S}{OURCE}&\multispan2 \hfil $R_F^a$\hfil
&\multispan2 \hfil $R_F^b$\hfil&References&$t^c$&\mdot$_{\rm CO}$&
$L_{FIR}$&&References&
$\nu$&$S_\nu$&&References&\quad log $N_L^d$&\multispan2\hfil\quad\mion\cr
&&&&&&(yr)&(10$^{-5}$ \msunyr)&($L_\odot$)&&
&(GHz)&(mJy)&&&(s$^{-1}$)&\multispan2\hfil\quad(10$^{-5}$ \msunyr)\cr
\noalign{\vskip 2pt \hrule \vskip 2pt}
\cr
\multispan{18} \hfil A. Our Observations \hfil \cr
\noalign{\vskip 2pt \hrule \vskip 2pt}
W3 IRS 5 &0&.019    &0&.15    &1&3100& 4.8&200,000&    & 2
         &15.4&   7.0&    & 3&45.6&  1&.1 \cr
GL 490   &0&.10     &0&.82    &1&2300& 1.3&   1700&    & 4
         &23.1&   3.2&    & 5&44.6&  0&.14\cr
NGC 2071 &0&.082    &0&.66    &1&3200& 0.3&    520&    & 6
         &15.0&   9.2&$^e$& 7&44.2&  0&.09\cr
W28 A2   &0&.046$^f$&0&.36$^f$&1&1900&56.4&420,000&    & 8
         &15.0&6540.5&    & 9&48.8&280&.  \cr
GL 2591  &0&.0085   &0&.068   &1&2000& 6.8& 90,000&    &10
         & 5.0&  79.4&$^g$&11&46.5&  9&.1 \cr
S140     &0&.015    &0&.12    &1&4200& 5.2&   5000&$^h$&12
         &15.0&   7.7&$^h$&13&44.8&  0&.29\cr
Cepheus A&0&.041    &0&.32    &1&1600& 2.8& 25,000&    &14
         &23.1&  13.2&$^i$& 5&44.9&  0&.26\cr
Average  &0&.045 &0&.36 &   &2600\cr
\noalign{\vskip 2pt \hrule \vskip 2pt}
\cr
\multispan{18} \hfil B. Previous Observations \hfil \cr
\noalign{\vskip 2pt \hrule \vskip 2pt}
HH 7--11   &1&.2$^j$ &9&.8$^j$&15&\na&\na&      115&$^l$&16
           &5.0&0.8&$^m$&7&43.0&0&.02\cr
HH 7--11   &0&.65$^k$&5&.2$^k$&15\cr
HH 7--11   &0&.33    &2&.7    &17\cr
IRAS 03282+3035&0&.72$^j$&5&.7$^j$&18&\na&\na&$\leq$1.5&    &18
           &   &   &    & &\na &\multispan2\hfil\na\cr
IRAS 03282+3035&0&.23$^k$&1&.8$^k$&18\cr
NGC 2071   &0&.24    &1&.9    &19\cr
NGC 7538 IRS 9&0&.62$^n$&4&.9$^n$&20&\na&\na&    40,000&    &20
           &   &   &    & &\na &\multispan2\hfil\na\cr
NGC 7538 IRS 9&0&.18$^o$&1&.5$^o$&20\cr
\noalign{\vskip 2pt \hrule \vskip 10pt}}}
\setbox2=\vbox{\tablesize\hsize=\wd1\goodfil
$^a$ Ratio of driving forces
     assuming the EHV outflow is the stellar wind ([C]/[CO] = 1). \par
$^b$ Ratio of driving forces
     assuming the EHV outflow is swept-up molecular gas ([C]/[CO] = 8). \par
$^c$ \hbox to 9cm{Characteristic timescale of the EHV outflow. \hfil}
$^d$ Rate of production of Lyman-continuum photons. \par
$^e$ \hbox to 9cm{From NGC 2071 IRS 1 alone. \hfil}
$^f$ From blue outflows only. \par
$^g$ \hbox to 9cm{From GL 2591 (1) alone. \hfil}
$^h$ From S140 IRS 1 alone. \par
$^i$ From Cepheus A (2) alone. \par
$^j$ \hbox to 9cm{Assumes IHV outflow is part of EHV outflow. \hfil}
$^k$ Assumes IHV outflow is part of HV outflow. \par
$^l$ \hbox to 9cm{From SVS 13 alone. \hfil}
$^m$ From NGC 1333 (5) alone. \par
$^n$ \hbox to 9cm{From red EHV and red HV outflows only. \hfil}
$^o$ From blue HV, red EHV, and red HV. \par
\refs (1) This paper; (2) Werner et al. 1980; (3) Colley 1980;
(4) Harvey et al. 1979 (corrected according to our distance);
(5)~Simon et~al. 1983; (6) Butner et al. 1990;
(7)~Snell \& Bally 1986; (8) Emerson, Jennings, \& Moorwood 1973;
(9) Wood \& Churchwell 1989; (10) Lada et~al. 1984; (11)~Campbell 1984;
(12) Lester et~al. 1986; (13) Evans et al. 1989;
(14) Koppenaal et~al. 1979 and Evans et~al. 1981; (15) Koo 1990;
(16)~Cohen \& Schwartz 1987; (17) Bachiller \& Cernicharo 1990;
(18) Bachiller et al. 1991; (19)~Chernin \& Masson 1992;
(20) Mitchell \& Hasegawa 1991.}
\setbox3=\vbox{{\box1}{\box2}}
\rotl{3}\vfill\eject
}

\vfill\eject

\clearpage
\centerline{\bf FIGURE CAPTIONS}

\figcap{1}{({\it a\/}) CO spectra of W3 IRS 5 with baselines.
The upper four panels show the full spectra of the lines,
and the middle and lower four panels have blown-up vertical scales
to show the details of the wings.
All the horizontal axes cover the same velocity extent.
Vertical dashed lines below the baselines are the wing boundaries
(also given in Table~4).
({\it b\/}) Contour maps of the integrated intensity
for each wing of W3 IRS 5 in the CO \jj32\ line.
The lower panel shows the maps of the inner wings,
and the upper panel shows those of the outer wings.
Solid contours are for red wings, and dashed contours for blue wings.
At the upper right corner of each panel are
the integrated antenna temperature ($\int T^*_A dV$)
of the lowest contour level, the highest contour level,
and the spacing between contours in K {km~s$^{-1}$}.
Maps were made with a 12{${''}$}\ grid.
Maps also include markers at the positions of IRS~6, IRS~5, IRS~7, and IRS~3
($\bullet$) from left to right (Wynn-Williams, Becklin, \& Neugebauer 1972),
and H$_2$O maser sources ($\triangle$) from Forster, Welch, \& Wright (1977).
The maps of the outer wings show a compact source centered at or near IRS~5.
The emission in the red wing appears slightly extended toward the southwest
(cf. Mitchell et~al. 1991; Claussen et al. 1984).
({\it c\/}) At the central position of W3 IRS 5,
the line ratio ($T_R$) of CO \jj32\ to \jj21\ in the top panel,
CO \jj32\ to $^{13}$CO \jj32\ in the second panel,
optical depth of CO \jj32\ line in the third panel,
and line ratio of CO \jj21\ to $^{13}$CO \jj21\ in the bottom panel.
Values for line ratios have been left out whenever $T_R$'s drop
below the thresholds given in Table~3.
Vertical dashed lines are the wing boundaries.
Solid exponential curves in the inner wings in the third panel are
the least-squares fits of $\tau_{32}$.
For the outer wings, we assumed $\tau_{32} \ll$ 1.
The line ratios \rco, \rhi, and \rlo\ rise toward the line wings
until they become undefined because the line in the divisor is too weak.
The ratio \rco\ implies low optical depth, even for the inner wings.
Even assuming the maximum ratio for the beam filling factors,
\rco\ becomes large enough to require \hbox{$\tau_{32} < 1$}
for inner and outer wings, contrary to the $\tau_{32}$ calculated from \rhi .
In fact, \rco\ is illegally high [$> (3/2)^4$] at some channels in the outer
wings, perhaps because of a pointing error in the CO \jj21\ data
(note that the blue EHV outflow of W3 IRS 5 is
the most compact among the sources in this paper).
Nonetheless, we interpret the high ratios as strong evidence for optically
thin emission.}

\figcap{2}{({\it a\/}) The same as Fig. 1{\it a\/} for GL 490.
Only GL~490 has CO \jj21\ wings as strong as the CO \jj32\ wings.
The \jj10\ line is also comparable in strength to the \jj21\ line
over the velocity interval of the inner wings (Margulis \& Lada 1985).
({\it b\/}) The same as Fig. 1{\it b\/} for GL 490.
Maps were made with a 9{${''}$}\ grid.
For the maps of red wings the lowest contours are the southernmost ones.
Markers indicate the position of an infrared source ($\bullet$)
from McGregor, Persson, \& Cohen (1984).
The radio continuum source is coincident
with the infrared source.
Our map does not cover the whole outflow region and does not
reveal a clear morphology.
({\it c\/}) The same as Fig. 1{\it c\/} for GL 490.
The ratio \rco\ is $\sim$0.9 throughout the line core and the wings,
suggesting that the gas is cold
($T_{ex}$ {$\rlap{\raise.4ex\hbox{$<$}}\lower.55ex\hbox{$\sim$}\,$}\ 50 K)
and/or the wings are optically thick,
even at extremely high velocities.
So we assumed that $\tau_{32}$ is 2 in the EHV wings.}

\figcap{3}{({\it a\/}) The same as Fig. 1{\it a\/} for NGC 2071.
There is a narrow ($\Delta V \approx$ 20 {km~s$^{-1}$}) spectral feature
at $\sim$55 {km~s$^{-1}$}\ near the southwestern edge of our map,
consistent with the feature found by Chernin \& Masson (1992).
({\it b\/}) The same as Fig. 1{\it b\/} for NGC 2071.
Maps were made with a 9{${''}$}\ grid.
Markers indicate IRS~4, IRS~2, IRS 3, and IRS 1 ($\bullet$)
from northeast to southwest (Persson et~al. 1981)
and an H$_2$O maser source ($\triangle$) from Genzel \& Downes (1979).
The bipolar nature of the outflow is clear
although our map does not completely cover the blue and red peaks
(cf. Bally 1982).
Chernin \& Masson (1992) have mapped the EHV gas to larger distance
and found that it covers a smaller region than the HV gas.
({\it c\/}) The same as Fig. 1{\it c\/} for NGC 2071.
The isotopic ratio, \rlo, in the inner wings agrees well
with that of Snell et al. (1984).
The interpretation of the line ratios is uncertain for this source
because none of the wings have a peak at the central position.
Both \rhi\ and \rlo\ show the typical rise toward the outer wings.
The red side of \rhi\ and the blue side of \rlo\ seem to decrease
beyond $|V-V_0| \approx$ 15 {km~s$^{-1}$},
but the uncertainties are very large.
The ratio \rco\ rises only to the red side of the spectrum.}

\figcap{4}{({\it a\/}) The same as Fig. 1{\it a\/} for W28~A2.
The peak at $V = 100$ {km~s$^{-1}$}\ in the CO \jj32\ spectrum is
the CS \jj76\ line in the other sideband.
Note that even the CS line is as extended as our inner wings,
which suggests that the inner wings are emitted by dense gas
(Plume, Jaffe, \& Evans 1992).
The baseline of the CO \jj32\ line is quite poor
because of the broad CS line.
Since the inner red wing of CO \jj32\ contains a dip
which may be either emission in the reference beam or self-absorption,
we define the inner red wing
using the $^{13}$CO \jj32\ line.
({\it b\/}) The same as Fig. 1{\it b\/} for W28~A2.
Maps were made with a 12{${''}$}\ grid.
The map of the inner red wing is omitted
because the spectra contain a dip.
Markers indicate H$_2$O maser sources ($\triangle$)
from Genzel \& Downes (1977),
and a compact \hii\ region ($\times$) from Wood \& Churchwell (1989).
The contour map shows a compact source
with both blue and red wings peaking at the position of an infrared
source and H$_2$O maser and also
well within the extent of the radio continuum emission.
({\it c\/}) The same as Fig. 1{\it c\/} for W28~A2.
The line ratios in the outer wings are somewhat unreliable
because of poor baselines.
The ratios \rlo\ and \rhi\ rise toward the wings
until they become undefined,
which indicates that the optical depth is getting smaller
as the velocity increases.}

\figcap{5}{({\it a\/}) The same as Fig. 1{\it a\/} for GL 2591.
The peak at $V= 90$ {km~s$^{-1}$}\ is the CS \jj76\ line.
({\it b\/}) The same as Fig. 1{\it b\/} for GL 2591.
Maps were made with an 18{${''}$}\ grid.
Markers indicate an infrared source ($\bullet$) from Lada et al. (1984),
an H$_2$O maser source ($\triangle$) from Wynn-Williams et al. (1977),
and radio continuum peaks ($\times$) from Campbell (1984).
The integrated intensities of the inner wings
do not peak within our map.
The CO \jj21\ map of Mitchell et al. (1991) shows
that the peaks of the inner wings are indeed beyond the boundaries of our map.
Both outer wings clearly have compact emission peaked at the same position.
In this region, the H$_2$O maser source, the infrared source,
and the strongest radio continuum source are
all displaced from each other, lying along a NE-SW line.
The peak of the outer wing emission is displaced from all these sources,
but is closest to the radio continuum source.
The offset from all the sources is smaller than the beam size,
but larger than the combined pointing uncertainties
and positional uncertainties of the maser and infrared source.
If observations with higher resolution confirm this offset,
it would suggest that the source of radio continuum
emission, rather than the infrared source, drives the outflow.
({\it c\/}) The same as Fig.~1{\it c\/} for GL 2591.
The ratios \rhi\ and \rco\ have minima at the line center
and rise toward the line wings until they become undefined,
suggesting that the outer wings are optically thin.
Since only one channel in the inner red wing has a well-defined \rhi ,
we included the outermost channel of the line core
to make the $\tau_{32}$ fit.}

\figcap{6}{({\it a\/}) The same as Fig. 1{\it a\/} for S140.
The CO \jj32\ spectrum shows two CS \jj76\ lines
(at $V = -105$ and 90 {km~s$^{-1}$})
because it is a sum of data taken during two periods when the source
had two different topocentric velocities.
({\it b\/}) The same as Fig. 1{\it b\/} for S140.
Maps were made with a 12{${''}$}\ grid.
Markers indicate infrared and radio sources,
IRS 2, IRS 1, and IRS 3 ($\bullet$) clockwise from the top
(Hackwell, Grasdalen, \& Gehrz 1982; Evans et al. 1989), and
an H$_2$O maser source ($\triangle$) from Genzel \& Downes (1979).
The inner wings show a clear bipolar pattern (cf. Hayashi et~al. 1987).
The contour map of the outer wings looks clumpy
because of several secondary peaks.
S140 is the only source in our sample in which the outer wings appear clumpy.
({\it c\/}) The same as Fig. 1{\it c\/} for S140.
The isotopic ratio, \rlo, of the inner wings is
consistent with Snell et al. (1984).
Both isotopic ratios rise toward the outer wings.
The ratio \rco\ rises toward the blue wing,
but decreases toward the red wing,
perhaps a result of observing with different beam sizes
a wing whose spatial distribution does not peak at the center of the source;
i.e., the CO \jj21\ beam is larger
than the CO \jj32\ beam
and contains more of the red peak.
We assumed $\tau_{32}=2$ in the red wing.}

\figcap{7}{({\it a\/}) The same as Fig. 1{\it a\/} for Cepheus A.
The peak at $V = -110$ {km~s$^{-1}$}\ is CS \jj76\ line.
({\it b\/}) The same as Fig. 1{\it b\/} for Cepheus A.
Maps were made with a 12{${''}$}\ grid.
Markers indicate a mid-infrared source ($\bullet$)
from Beichman, Becklin, \& Wynn-Williams (1979),
radio component 2 of Hughes (1985) ($\times$),
and an H$_2$O maser source ($\triangle$) from Lada et~al. (1981).
The contour maps clearly show the bipolar nature of the inner wings.
The outer wings, however, show a rather complex structure.
The outer blue wing has a local peak at the center position,
while the red wing has a displaced peak.
The peak of the blue wing coincides within the uncertainties
with an H$_2$O maser source and a radio continuum source.
({\it c\/}) The same as Fig. 1{\it c\/} for Cepheus A.
The ratios \rhi\ and \rlo\ have minima at the line center
and rise toward the line wings until the denominators become
less than the thresholds.
Since \rco\ is also consistent with low $\tau_{32}$,
we have assumed optically thin emission in the outer wings.}

\figcap{8}{$D(n, T_K)$ (defined by eq. [6]) is plotted
for reasonable values of $n$ (10$^4$--10$^6$ cm$^{-3}$)
and $T_K$ (10--200~K),
as calculated from the LVG model, and from the LTE assumption.
In the LVG calculations,
we assumed a constant CO column density per unit velocity interval,
$N/dV$ = 10$^{16}$ cm$^{-2}$ km$^{-1}$ s,
typical of our derived values,
but $D(n, T_K)$ is insensitive to $N/dV$.}

\figcap{9}{The ratio of EHV to HV driving force
versus luminosity ({\it upper panel\/})
and the rate of production of Lyman-continuum photons ({\it lower panel\/}).
We assumed that the EHV gas is the stellar wind ([C]/[CO] = 1)
and that the HV outflow is swept-up molecular gas \hbox{([C]/[CO] = 8)}.
Filled pointers are from this paper,
and open pointers are from references given in Table~6.
If there are multiple references for one source,
corresponding pointers are connected by vertical lines.
See Table 6 for details.}

\figcap{10}{\rco\ as a function of $\tau_{32}$ for various $T_{ex}$.
It is assumed that $\tau_{21}$ follows eq.~(4).
Numbers on the left vertical axis are for $\phi$ = (3/2)$^2$ (point source),
and those on the right vertical axis are for $\phi$ = 1.0 (extended source).}

\figcap{11}{The observed excitation temperature, $T_{ex,obs}$,
as a function of the real excitation temperature,
$T_{ex}$, with two components in the beam.
It is assumed that \hbox{$X$ = 56}, \hbox{$\tau _{u{\rm A}}$ = 1},
$\tau _{u{\rm B}}$ = 10, $T_{ex\rm A}$ = $T_{ex\rm B}$,
and all the filling factors are the same.
The solid curve is for the combination of
CO \jj21\ and CO \jj10,
and the dashed curve is for the combination
of CO \jj32\ and CO \jj21\ transitions.
The straight dotted line is $T_{ex,obs}$ = $T_{ex}$.}

\end{document}